\documentclass[a4paper,11pt]{article}

\usepackage[hmargin=2cm,vmargin=2.5cm]{geometry,}
\usepackage{graphicx, amssymb, textcomp, hyperref, amsmath}
\usepackage{epsfig}
\numberwithin{equation}{section}

\title{\bf Quantum superintegrable system with a novel chain structure of quadratic algebras}
\author{\bf Yidong Liao, Ian Marquette and Yao-Zhong Zhang\,\footnote{Corresponding Author}}
\date{\it\small School of Mathematics and Physics, The University of Queensland, Brisbane 4072, Australia\\ 
{}
{\tt\small Email: yidong.liao@uq.net.au, i.marquette@uq.edu.au, yzz@maths.uq.edu.au}}

\begin{document}

\maketitle

\vspace{0.5in}
\noindent {\bf Abstract:} We analyse the $n$-dimensional superintegrable Kepler-Coulomb system with non-central terms. We find a novel underlying chain structure of quadratic algebras formed by the integrals of motion. We identify the elements for each sub-structure and obtain the algebra relations satisfied by them and the corresponding Casimir operators. These quadratic sub-algebras are realized in terms of a chain of deformed oscillators with factorized structure functions. We construct the finite-dimensional unitary representations of the deformed oscillators, and give an algebraic derivation of the energy spectrum of the superintegrable system.

\vspace{.3in}
%\noindent{\em PACS numbers:} 03.65.-w, 03.65.Fd, 03.65.Ge, 02.30.Ik

%\noindent{\em Keywords:} 

\section{Introduction}
Superintegrable systems are a distinct subclass of integrable systems which possess more integrals of motion than the degrees of freedom and can be solved analytically and algebraically. Famous examples of superintegrable systems include the classical Kepler system and the Hydrogen atom. Classification of two-dimensional quadratically superintegrable systems on 2D conformally flat spaces was given,  and the role of the underlying quadratic algebras with three generators  as well as the connection with the full Askey Scheme of orthogonal polynomials were revealed \cite{review}. The classification was extended to three dimensions, but the case of degenerate systems has not been completed. Some quadratic algebras have been constructed from their integrals of motion and the role of linearly independent integrals was highlighted \cite{miller1, miller2, miller3, escobar1, Daskaloyannis}. However, a systematic construction of symmetry algebras, Casimir operators and representation theory is still lacking. It is conjectured that symmetry algebras of $n$-dimensional superintegrable systems would be in general higher rank polynomial algebras, but only a few isolated examples of such algebras are known\cite{Fazlul,Fazlul2,Fazlul3,vinet1,lliev1}.

Approaches to obtain algebraic derivation of spectra for quantum systems with polynomial symmetry algebras involving three generators were proposed in \cite{deformed,Da,de}. These methods are based on the construction of Casimir operators and realizations of the polynomial algebras as deformed oscillator algebras.  However, the case corresponding to higher rank polynomial algebras remains largely unexplored. This is due to the fact that it is quite involved to construct such algebraic structures and their corresponding Casimir operators. Recently, it was shown \cite{Fazlul} that the symmetry algebra of a family of $d$-dimensional Kepler-Coulomb systems with non-central terms is the direct sum of the quadratic algebra $Q(3)$ and the Lie algebra $so(d-1)$ i.e. $Q(3) \oplus so(d-1)$. This algebra structure was exploited to obtain the algebraic derivation of the spectra of the models.

Rodriguez and Winternitz \cite{model}, and independently Kalnins et al
\cite{Kalnins02}, proposed an $n$-dimensional generalization of the 3D superintegrable Kepler-Coulomb Hamiltonian with non-central terms considered in \cite{Evans90,Evans91,Grosche95,Kalnins99}. This new $n$-dimensional system is maximally superintegrable and has quadratic integrals of motion. It was solved in \cite{model} in the parabolic and spherical coordinates using the separation of variables and its wave functions were given in terms of orthogonal polynomials. However, the hidden symmetry algebra for this model was unknown.

In this paper we will build from the integrals of motion the algebraic structure underlying the degeneracies of this system and give an algebraic derivation of its energy spectrum. It will be shown that the symmetry algebra of this model consists of multiple overlapping quadratic algebras. By overlapping we mean there are common elements among the multiple quadratic algebraic structures, more precisely, there exist a chain of quadratic algebras, each of which shares several common elements with its two neighbours. 
We examine the overlapping quadratic algebras of the system and identify the elements of each sub-structure. Through analysing the results of the cases of specific dimensions, we propose and verify the general formula of the quadratic algebra for the $n$-dimensional system. Utilizing the common elements between the algebras, as well as the maximal set of commuting element, we perform an algebraic derivation for the energy spectrum of the system.

The structure of the paper is as follows: In Section 2, the superintegrable Hamiltonian system in $n$-dimensional Euclidean space and its analytical solution of the energy spectrum will be presented. In Section 3, we construct the quadratic algebras and their Casimir operators. In section 4, we obtain the realization of the quadratic algebra in terms of the deformed oscillator algebras and the corresponding structure functions. In section 5, we give an algebraic derivation of the energy spectrum of the system. Finally, in Section 6, we present some discussions with remarks on the physical and mathematical significance of these algebras.

\section{The $n$-dimensional quantum superintegrable system}

The Hamiltonian of the $n$-dimensional superintegrable system in \cite{model,Kalnins02} is given by 
\begin{equation}
H=-\frac{1}{2}\sum_{i=1}^{n}\frac{\partial^{2}}{{\partial x_{i}}^{2}}- \frac{\gamma}{r}+ \sum_{i=1}^{n-1}\frac{\beta_{i}}{x_{i}^{2}},\label{2.1}
\end{equation}
where $r=\sqrt{\sum_{i=1}^{n}{x_{i}^{2}}}$ and ${\beta_{i}},\gamma$ are real parameters.
It can be seen that this system contains the $n$-dimensional hydrogen atom as a special case.
This is a superintegrable model with the most general potential allowing the separation of variables in the same four coordinate systems as the hydrogen it self.

The authors in  \cite{model}  solved the corresponding Schr\"odinger equation in the parabolic and spherical coordinates using the separation of variables and obtained the energy spectrum,
\begin{equation}
E=-\frac{\gamma^2}{2(N_1+N_2+2\sum_{i=1}^{n-2}J_i+\sum_{i=1}^{n-1}
p_i+\frac{n-1}{2})^2},\label{2.2}
\end{equation}
where $N_1$ and $N_2$ are positive integers, $J_i$ are angular momentum quantum numbers and $p_i$ are related to ${\beta_{i}}$ via ${\beta_{i}}= \frac{1}{2}p_i(p_i-1)$. We see that the bound state energy is given by a shifted Balmer formula.

The aim of this paper is to derive the energy spectrum (\ref{2.2}) using algebraic method. It will be shown that the underlying algebraic structure of this system consists of multiple overlapping quadratic algebras.

\section{Overlapping quadratic algebras of the superintegrable system}

\subsection{Elements of the proper sub-structures }

The system (\ref{2.1}) possesses the following integrals of motion \cite{model}
\begin{equation}
X=\frac{1}{2}\sum_{k=1}^{n-1}(\hat{L}_{nk}\hat{p}_{k}+\hat{p}_{k}\hat{L}_{nk})+2x_{n}( \frac{\gamma}{2r}-\sum_{i=1}^{n-1}\frac{\beta_{i}}{x_{i}^{2}}),\label{3.1}
\end{equation}
\begin{equation}
Z_{l}=\sum_{1 \leq i < k \leq l+1}\hat{L}_{ik}^{2}-2 ( \sum_{i=1}^{l+1}x_{i}^{2}) ( \sum_{k=1}^{l+1}\frac{\beta_{k}}{x_{k}^{2}}), ~~~~ 
{  } 1 \leq l \leq n-2, \label{3.2}
\end{equation} 
\begin{equation}
Y_{p}=\sum_{p \leq i < k \leq n}\hat{L}_{ik}^{2}-2 ( \sum_{i=p}^nx_{i}^{2})( \sum_{k=p}^{n-1}\frac{\beta_{k}}{x_{k}^{2}}),~~~~ 
{   }  1 \leq p \leq n-1,\label{3.3}
\end{equation}
where $\hat{p}_{k}=\frac{\partial}{\partial x_k}$, $\hat{L}_{ik}=x_i\frac{\partial}{\partial x_k}-x_k\frac{\partial}{\partial x_i}$.
As pointed out in \cite{model}, these integrals obey the commutation relations, 
\begin{equation}
[X,Z_i]=0=[Z_i,Z_j],~~~~~ [Y_j,Y_i]=0=[Y_1,Z_i].\label{Cartan-type}
\end{equation}
By calculating the other commutation relations satisfied by these integrals explicitly for $n=3,4,5,6$, we are able to identify the proper sub-structures, each of which is generated by two non-commutating generators, along with several central elements.By analyzing the obtained results corresponding to the 3D, 4D, 5D and 6D cases, we can see the pattern of the proper sub-structures and thereby identify the elements of the proper sub-structures for the general 
$n$-dimensional case:
\begin{center}
 \begin{tabular}{||c c c  ||} 
 \hline
Sub-structure & Generators & Central elements \\ [0.3ex] 
 \hline
 1 &  $Y_1$,~$X$ & $H$, $Z_l~ (1\leq l\leq n-2)$\\  [0.5ex] 
%  \hline
% 2 &  $Y_2$,~$Z_1$ & $H$, $Y_1$, $Y_3$\\  [0.5ex] 
 \hline
 i~~($2\leq i\leq n-1$) &  $Y_i$,~$Z_{i-1}$ &  $H$, $Z_j~ (j<i-1)$, $Y_1$,  $Y_k ~(k>i)$ \\  [0.5ex] 
\hline
\end{tabular}
\end{center}

Note that the $n$-dimensional system (\ref{2.1}) comes from the generalization of the 3D Hamiltonian in \cite{Evans90,Evans91,Grosche95,Kalnins99}). The $n=2$ case needs a special consideration as indicated by the range of index $l$ in (\ref{3.2}). (See the remarks after (\ref{K1a}) in next section.) Furthermore, as seen from the range of index $p$ in (\ref{3.3}), for any fixed dimension value $n$, there is no integral $Y_n$. Therefore it should be understood {\em throughout this paper} that $Y_n\equiv 0$ for any given $n$ value.

\subsection{Quadratic sub-algebra structures}

In the previous sub-section we have identified the elements (generators and central elements) of each sub-structure. Using the expressions (\ref{3.1})-(\ref{3.3}), we now compute the commutation relations satisfied by the elements of each sub-structure and the corresponding Casimir operators. We find that the elements in each sub-structure form a quadratic algebra with central elements which are elements of  other sub-structures.

\vskip.2in
\noindent \underline{\bf\large Sub-structure $1$}: 
\vskip.1in
It can be shown that the  generators $ Y_ {1}, X$  satisfy the following commutation relations:
\begin{eqnarray}
[Y_ {1}, X] &=& C_ {1},\nonumber\\
{}[Y_ {1}, C_ {1}]& =& -2\{Y_ {1}, X\} + (n - 3) (n - 1) X,\nonumber\\
{}[X, C_ {1}] &= &
 2 X^{2} - 4 H Z_ {n - 2} + 8 Y_ {1} H - (n - 1)^2 H - 2 \gamma^{2}.
 \label{sub-structure1}
\end{eqnarray}
These relations are obtained by the method of undetermined coefficients, assuming the most generic expression of the quadratic function of the generators and central elements on the right hand side of the second and third equations above. It turns out that $H, Z_{n-2}$, which commute with each other, also commute with $ Y_ {1}, X$ and $C_1$. Thus $ Y_ {1}, X$ form a quadratic algebra with central elements  $H, Z_{n-2}$. 
The structure of this sub-algebra can be illustrated below in Figure 1 in which a line between the integrals means that the two elements commute with each other.
\begin{figure}[h]
    \centering
    \includegraphics[width=4cm]{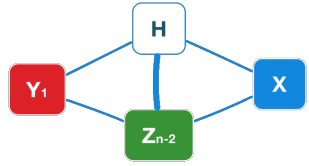}
    \caption{Sub-structure 1 with generators $ Y_ {1}, X$ and central elements $H, Z_{n-2}$}
    \label{fig:my_label-1}
\end{figure}

The Casimir operator $K_1$ of this quadratic algebra, written in terms of the generators and central elements, is found to be given by
\begin{small}
\begin{equation}
K_1 = C_1^2+2 \{Y_1,X^2 \}+2 Y_1  (-2 \gamma^2-H (n-1)^2-4 HZ_{n-2} )\\+8 H Y_1^2+(4-(n-3) (n-1)) X^2.\label{K1a}
\end{equation}
\end{small}
The Casimir operator is a cubic function of the integrals $ Y_ {1}, X$,  $H, Z_{n-2}$.

Some remarks are in order. 
(\ref{sub-structure1}) is the sub-structure  for a system with dimension $n\geq 3$. However, it is natural to expect that the $n=2$ model should appear as a particular case in the genuine $n$-dimensional system.  As seen from (\ref{3.1}-\ref{3.3}), when $n=2$ there are only two integrals $\{X, Y_1\}$. So to include the $n=2$ case in the relations (\ref{sub-structure1}), we need to define the value of $Z_0$. A direct computation using the integrals $\{X, Y_1\}$ given by (\ref{3.1}) and (\ref{3.3}) shows that when $n=2$, we have
\begin{eqnarray}
[Y_ {1}, X] &=& C_ {1},\nonumber\\
{}[Y_ {1}, C_ {1}]& =& -2\{Y_ {1}, X\} - X,\nonumber\\
{}[X, C_ {1}] &= &
 2 X^{2} - 4 H (-2\beta_1) + 8 Y_ {1} H -  H - 2 \gamma^{2},
 \label{n=2}    
\end{eqnarray}
which coincide with (\ref{sub-structure1}) with $n=2$ if one defines $Z_0$ to be $-2\beta_1$. So from now on (and {\em throughout the paper}), it should be understood that $Z_0\equiv -2\beta_1$.

\vskip.2in
\noindent \underline{\bf\large Sub-structure $i ~(2\leq i\leq n-1)$}:
\vskip.1in
It can be shown that the quadratic algebra formed by the generators $ Y_ {i}, Z_ {i-1}$ $(2\leq i\leq n-1)$ is given by the commutation relations:
\begin {eqnarray} 
[Z_{i-1},Y_i]&=&C_i,\nonumber\\
{}[Z_ {i-1}, C_ {i}] &=& -8Z_{i-1}^2- 8\{Z_ {i-1}, Y_ {i}\} 
   -4((i-2)n-(i^2-i-4)+4\beta_ {i} ) Z_ {i-1} \nonumber\\
& &+4(n-i+1)(n-i-3)Y_ {i} + 8 (Y_ 1  +  Z_{i-2} ) Z_ {i-1}-4(n-i-3+ 4\beta_ {i}) Y_1\nonumber\\
& &-4((n-i)(n-i-3)-4\beta_ {i})Z_ {i-2} +8(n-i+1)(n-4)\beta_ {i} \nonumber \\
& & +4(i-1)(n-i+1) Y_{i+1} - 8 Y_ 1  Y_ {i+1} +8   Z_ {i-1}Y_ {i+1}  + 8 Z_{i-2} Y_ {i+1}\nonumber\\
%  -4(i-4+4\beta_ {i})Y_ {1}+4i(i-4)Y_ i+16 (-1   
%  + \beta_ {i}) Y_{i+1}+4i(n-i) Z_ {i-2}\nonumber\\
%& &-4((i-2)n-(i^2-i-4)+4\beta_ {i}) Z_ {i-1}+8i(n-4)\beta_ {i}
%  -8Z_ {i-1}^2  - 8\{Z_ {i-1}, Y_ {i}\}\nonumber\\
%& & + 8 Y_ 1  Z_ {i-1} + 8 Y_ {i+1}  Z_ {i-1} -8 Z_ {i-2}  Y_ 1 + 8 Z_  {i-2}  Y_ {i+1}  + 8 Z_{i-2}   Z_ {i-1}, \nonumber\\
{}[Y_ {i},C_ {i}]&=& +8Y_ {i}^2  +8\{Z_ {i-1}, Y_ {i}\}-4i(i-4)Z_ {i-1}+4((i-2)n-(i^2-i-4)+4\beta_ {i}) Y_ {i}\nonumber\\
& &- 8 Z_{i-2}   Y_ {i}- 8 Y_ 1  Y_ {i}+4(i-4+4\beta_ {i})Y_ {1}+8 Z_ {i-2}  Y_ 1-4i(n-i) Z_ {i-2}\nonumber\\
& &-8i(n-4)\beta_ {i}+4 ((i-4)(i-1)-4 \beta_ {i}) Y_{i+1}   - 8 Z_  {i-2}  Y_ {i+1}  - 8 Y_ {i+1}  Y_ {i}.
%  - 4(n-i-3 + 4\beta_ {i}) Y_1
%   +4((i-2)n-(i^2-i-4)+4\beta_ {i} ) Y_ i \nonumber\\
%& &+4(i-1)(n-i+1) Y_{i+1}+4((n-i)(n-i-3)-4\beta_ {i})Z_ {i-2}\nonumber\\
%& &-4(n-i+1)(n-i-3)Z_ {i-1}  -8(n-i+1)(n-4)\beta_ {i}+8Y_i^2\nonumber \\
%& &+ 8\{Z_ {i-1}, Y_ {i}\} - 8 Y_ 1  Y_ i + 8 Y_ 1  Y_ {i+1} -8 Y_ {i+1}  Y_ i - 8 Z_{i-2}  Y_ i - 8 Z_{i-2} Y_ {i+1}.
\label{sub-structure-i}
\end {eqnarray}
We remind that as mentioned previously it should be understood that $Y_n\equiv 0$ in the above relations for any fixed value of $n$.
These generic quadratic algebra formulas for the sub-structure 
$i$ of the system are obtained by examining the pattern in the results derived for each sub-structure i with $2\leq i \leq n-1$ up to $n=9$. 
For a fixed value of $i$ and $n$, the same method of undetermined coefficients has been used, assuming the most generic expression of the quadratic function of the generators and central elements. 
For a fixed $i$ the elements $Y_ {1},Y_ {i+1}, Z_ {i-2}$, which commute with each other, also commute with $Z_ {i-1},Y_ {i}, C_ {i}$, and thus are central elements of the sub-structure $i$.
The structure of the quadratic sub-algebra for any fixed $2\leq i\leq n-1$ can be illustrated in Figure 3:
\begin{figure}[h]
    \centering
    \includegraphics[width=4cm]{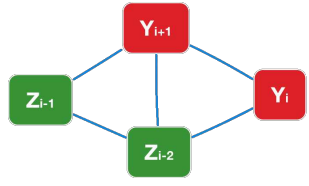}
    \caption{Sub-structure i, generators $ Y_ {i}, Z_ {i-1}$ and central elements  $Y_ {1},Y_ {i+1},Z_ {i-2}$ (we omit $Y_1$ in the graph above as $Y_1$ is the central element of all sub structure $i~(2\leq i\leq n-1)$ and note $Y_n\equiv 0$).}
    \label{fig:my_label-i}
\end{figure}

Similar to sub-structure 1, it can be seen that the Casimir operator $K_i$ for the sub-structure $i~ (2\leq i\leq n-1)$ is a cubic function of the integrals generators 
$ Y_ {i}, Z_ {i-1}$ $Y_ {1},Y_ {i+1}, Z_ {i-2}$. In terms of the generators and central elements, $K_i$ is given by
\begin{eqnarray}
K_{i}&=&  4(-i^2+4 \beta _i+(i-2) n+i+4 -2 Y_{i+1}-2 Z_{i-2}
   -2Y_1+16 ) \{Y_i,Z_{i-1} \}+8 \{Y_i,Z_{i-1}^2\}\nonumber\\
& &+8\{Y_i^2,Z_{i-1}\}+C_i^2  -32 Y_i  (i^2-4 \beta _i-(i-2) n-i-4
   +2 Y_{i+1}+2  Z_{i-2}+2 Y_1 )\nonumber  \\ 
& &-8Z_{i-1}  \left[  -4  (-i^2+4 \beta _i+(i-2)n+i+4 )+8 Y_{i+1}+8      
   Z_{i-2}+8 Y_1 +4 i (n-4) \beta _i\right.
   \nonumber\\
& &\left.+i (n-i)   Z_{i-2}
   +Y_1  (4-i-4 \beta _i )+ (4 \beta _i- (i-4) (i-1) )Y_{i+1}
   -2 Y_1 Z_{i-2}+2 Y_{i+1} Z_{i-2} \right]\nonumber\\
& &   +8 Y_i  \left[-2 (n-4) \beta _i
   (-i+n+1)+Y_1  (4 \beta _i-i+n-3 )- (i-1) (-i+n+1) Y_{i+1}\right.
   \nonumber\\
& &\left.+Z_{i-2} ((-i+n-3) (n-i)-4 \beta _i )
   -2 Y_{i+1} Z_{i-2}+2 Y_1 Y_{i+1} \right]\nonumber\\
& &-4   (-i+n-3) (-i+n+1) Y_i^2+64 Y_i^2  +4(16-i(i-4) ) Z_{i-1}^2.
  \label{Kia}
\end{eqnarray} 

The commutation relations (\ref{sub-structure1}) and (\ref{sub-structure-i}) form a subalgebra of the full quadratic algebra structure of the $n$-dimensional 
superintegrable model (\ref{2.1}). 

\subsection{Chain structure of the full quadratic algebra symmetry}

In the remaining of this section, we illustrate the chain quadratic algebra structure of the model. Part of the chain structure of the full quadratic algebra can be illustrated in the graph 3 below:

\begin{figure}[h]
    \centering
    \includegraphics[width=8cm]{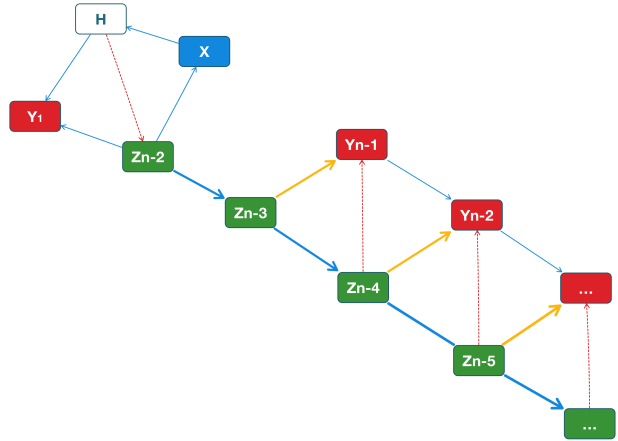}
    \label{fig:my_label-a}
\caption{As indicated in this graph, $Z_{n-2}$ is one of the central elements of sub-structure 1 and at same time, one of the generators of sub-structure $n-2$; $Z_{n-3}$ is one of the central elements of sub-structure $n-2$ and at same time, one of the generators of sub-structure $n-3$; $Y_{n-1}$ is one of the generators of sub-structure $n-2$ and at same time, one of the central elements of sub-structure $n-3$, and so on.}
\end{figure}

The full chain structure of the quadratic algebras is shown in the following graph 4:

\begin{figure}[h]
    \centering
    \includegraphics[width=8cm]{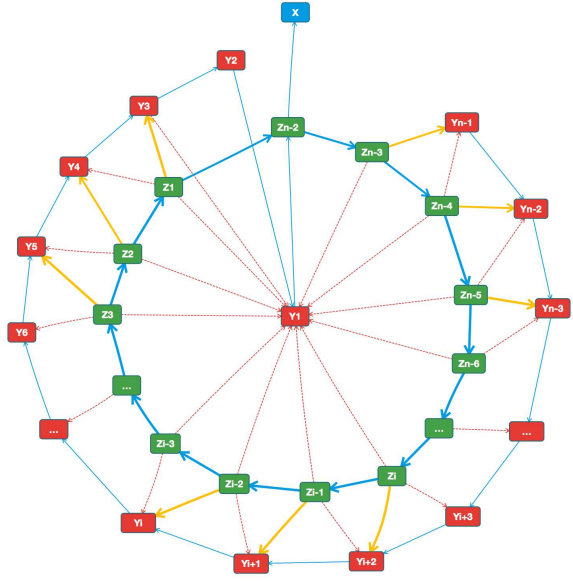}
    \label{fig:my_label-b}
\caption{Here for simplicity we have omitted $H$ in this graph but it should be noted that $H$ is one of the central elements of sub-structure 1 and it commutes with all the other integrals.
Note that the element $Y_1$ and those in the inner ring, $\{Y_1,Z_i|i=1,2,....n-2\}$, form a maximal commuting set, i.e. Cartan type subalgebra of the quadratic algebra because $[Z_i,Z_j]=[Y_1,Z_i]=0$ as seen from (\ref{Cartan-type}).}
\end{figure}

\section{Realization of each sub-quadratic algebra}

\subsection{Realization of sub-structure 1}

The quadratic algebra generated by $Y_1$ and $X$ can be realized by means of the deformed oscillator
\(\{b_1^\dagger, b_1, \mathcal{N}_1\}\), which satisfy \cite{deformed}
\begin{small}
\begin{equation}
 [ {\mathcal{N}_1}, b_1^\dagger  ] = b_1^\dagger, \quad  [ {{\mathcal{N}}_1}, b_1] = -b_1, \quad b_1^\dagger b_1 = \Phi_1({\mathcal{N}}_1 ),
\quad b_1 b_1^\dagger = \Phi_1({\mathcal{N}}_1+1 )
\label{oscillator-1}
\end{equation}
\end{small}

\noindent where $\Phi_1({\mathcal{N}}_1 )$ is certain structure function. This is seen as follow:

Set
\begin{eqnarray}
Y_1&=&-\left(\mathcal{N}_1+u_1\right){}^2+\frac{1}{4}
  +\frac{1}{4} (n-3) (n-1),\nonumber\\
X&=&{b_1^\dagger}\frac{1}{3145728 (u_1+{\mathcal{N}}_1)(u_1+{\mathcal{N}}_1+1)(2 (u_1+{\mathcal{N}}_1)+1)^2}\nonumber\\
& &~~+\frac{1}{3145728 (u_1+{\mathcal{N}}_1)(u_1+{\mathcal{N}}_1+1)(2 (u_1+{\mathcal{N}}_1)+1)^2}{b_1},
\end{eqnarray}
where $u_1$ is a free parameter to be determined later.Then we can show that $Y_1$ and $X$ defined above satisfy the commutation relations (\ref{sub-structure1}) provided that the structure function $\Phi_1(\mathcal {N} _ 1, u_1)$ in (\ref{oscillator-1}) has the form
\begin{eqnarray}
\Phi_ 1 (\mathcal {N} _ 1,u_1) &=& -32768   \left [12 (\mathcal {N} _ 1
  + u_1)^2 - 12 (\mathcal {N} _ 1 + u_1) - 1 \right] 
  \left[2 (\mathcal {N} _ 1 + u_1) - 1\right]^2\nonumber\\
& & ~~~\times[2 H (n - 3) (n - 1)+\mathcal {T}]
   - 32768 [2 (\mathcal {N} _ 1 + u_1) - 1]^2\nonumber\\
& &~~~\times\left[    -3 H (n - 3)^2 (n - 1)^2 + 8 H (n - 3) (n - 1) - 3 (n - 3) (n - 1)\mathcal {T}  + 4\mathcal {T}\right]\nonumber\\
& & +  98304 H \left[2 (\mathcal {N} _ 1 + u_1) - 3\right]
  [2 (\mathcal {N} _ 1 + u_1) + 1]
  \left[2 (\mathcal {N} _ 1 + u_1) - 1\right]^4  \nonumber\\
& & -  196608 K_ 1 (2 (\mathcal {N} _ 1 + u_1) - 1)^2\label{phi-1}
\end{eqnarray}
in which    $\mathcal{T}=  -2 {\gamma}^2-H (n-1)^2-4 {HZ}_{n-2}$.

In terms of the differential realisation (\ref{3.1})-(\ref{3.3}) for the integrals, the Casimir operator (\ref{K1a}), can be expressed in terms of the central elements $Z_{n-2}$ and $H$ of sub-structure 1 as 
\begin{equation}
K_1=-4 {\gamma}^2 Z_{n-2}-2{\gamma}^2 (n-3)-2 (n-3) (n-1)HZ_{n-2}-H (n-3) (n-1)^2\label{K1b}
\end{equation}   
This is obtained after lengthy computation by means of the method of undetermined coefficients, assuming the most generic cubic function of the central elements. It turns out that for this sub-algebra, only linear and quadratic terms appear in the expression of the Casimir operator.

Substituting (\ref{K1b}) into (\ref{phi-1}), we find that $\Phi_1(\mathcal {N} _ 1,H, u_1)$ can then be factorized as
\begin{eqnarray}
 \Phi_1(\mathcal {N} _ 1,H, u_1) &=&
  \left(\mathcal{N}_1 + u_1 - (\frac{1}{2}+m_{n-2})\right) \left(\mathcal{N}_1 + u_1 - (\frac{1}{2}-m_{n-2})\right)
  \left(\mathcal{N}_1 + u_1 - \frac{1}{2}\right)^2\nonumber\\
& &\left(\mathcal{N}_1 + u_1 - \left(\frac{1}{2}-\frac{i \gamma}{\sqrt{2H}}\right)\right)\left(\mathcal{N}_1 + u_1 - \left(\frac{1}{2}+\frac{i \gamma}{\sqrt{2H}}\right)\right),
\end{eqnarray}
where $m_{n-2}$ satisfies: 
\begin{equation}
Z_{n-2}=\frac{(n-3)^2}{4}-m_{n-2}^2.
\end{equation}

\subsection{Realization of sub-structure $i~ (2\leq i\leq n-1)$ }

We now come to the realisation of the sub-structure $i$ $(2\leq i\leq n-1)$. For fixed i, the corresponding quadratic algebra with generators \( Z_{i-1}, Y_i \) can be realized by means of the deformed oscillator
$ \{ b_i^\dagger, b_i, {\mathcal{N}}_i \}$,defined by \cite{deformed}
\begin{equation}
 [ {{\mathcal{N}}_i}, b_i^\dagger  ] = b_i^\dagger, \quad  [\mathcal{N}_i, b_i] = -b_i, \quad b_i^\dagger b_i = \Phi_i({\mathcal{N}}_i ),
\quad b_i b_i^\dagger = \Phi_i({\mathcal{N}}_i+1 ),\label{oscillator-i}
\end{equation}
where $\Phi_i$ is the structure function to be determined. We make the mapping
\begin{eqnarray}
Z_{i-1}&=&-4 \left(\left(\mathcal{N}_i+u_i\right){}^2-\frac{1}{16} (i-2)^2\right),\nonumber\\
Y_{i} &=&b_i^\dagger \frac{1}{206158430208 (u_i+\mathcal{N}_i) (u_i+\mathcal{N}_i+1) (2 (u_i+\mathcal{N}_i)+1)^2}\nonumber\\
& &+ \frac{1}{206158430208 (u_i+\mathcal{N}_i) (u_i+\mathcal{N}_i+1) (2 (u_i+\mathcal{N}_i)+1)^2}b_i\nonumber\\
& &+\frac{1}{4} \left(-4 \beta _i-(i-2) n+3i-4+2 Y_{i+1}
  +2 Z_{i-2}+2 Y_1\right)+2 \left((u_i+\mathcal{N}_i)^2-\frac{1}{4}\right)\nonumber\\
& &-\frac{1}{128 \left((u_i+\mathcal{N}_i)^2-\frac{1}{4}\right)}
   \left[ i(i-4)  \left(i^2+2i-8 \beta _i-2(i-2) n-8\right.\right.\nonumber\\
& &~~\left. +4 Y_{i+1}  +4 Z_{i-2}+4 Y_1\right)
+8 \left(2 i (n-4) \beta _i+ i (n-i) Z_{i-2}+Y_1 (4-i-4 \beta_i)\right.\nonumber\\
& &~~\left.\left.+(4 \beta _i- (i-4) (i-1)) Y_{i+1}-2 Y_1 Z_{i-2}+2 Y_{i+1}   Z_{i-2}\right)\right] 
\end{eqnarray}
where $u_i$ is certain free parameter to be determined in the next section. Then we can show that $Z_{i-1}$ and  $Y_{i}$ defined above satisfy the commutation relations in sub-structure 
$i~(2\leq i\leq n-1)$ with the structure function $\Phi_i(\mathcal{N}_i,u_i)$ given in terms of the Casimir operator $K_i$ and the central elements as
\begin{eqnarray}
\Phi_i (\mathcal{N}_i,u_i)&=& -805306368 K_i (2
    (\mathcal{N}_i+u_i)-1)^2\nonumber\\
& &+ 16777216 \left(12  (\mathcal{N}_i+u_i)^2-12  (\mathcal{N}_i+u_i)-1\right) (2  (\mathcal{N}_i+u_i)-1)^2\nonumber\\
& &~~\left[24(i-4)^2 i^2+32 (i-4) i (-i+n-3) (-i+n+1)-12 (i-4) i R
\right.\nonumber\\
& &~~\left.-16(2 \mathcal{P}+ \mathcal{Q})+R^2\right]
+3145728 \left(-2 (i-4)^2 i^2+ (i-4) iR+4 \mathcal{Q}\right)^2
\nonumber\\
& &-8388608 (2  (\mathcal{N}_i+u_i)-1)^2 
   \left(24 (i-4)^3   i^3\right.\nonumber\\
& &~~+24 (i-4)^2 i^2 (-i+n-3) (-i+n+1)-18 (i-4)^2 i^2 R\nonumber\\
& &~~-48   (i-4) i\mathcal{Q}-256 (i-4) i (-i+n-3) (-i+n+1)
  -48 (i-4) i   \mathcal{P}\nonumber\\
& &~~\left.+3 (i-4) i R^2+256 (\mathcal{P}- \mathcal{Q})+16 R^2+12 R\mathcal{Q}\right)\nonumber\\
& &-805306368(2  (\mathcal{N}_i+u_i)-3) (2  (\mathcal{N}_i+u_i)+1) (2  (\mathcal{N}_i+u_i)-1)^4 \nonumber\\
& &~~\times  (4 (-i+n-3) (-i+n+1)+4 (i-4) i- R)\nonumber\\
& &+3221225472 (2  (\mathcal{N}_i+u_i)-3)^2 (2  (\mathcal{N}_i+u_i)+1)^2 (2  (\mathcal{N}_i+u_i)-1)^4   ,\label{phi-i}
\end{eqnarray}  
where
\begin{eqnarray*}
\mathcal{P} &=& -8 (n-4) \beta _i (-i+n+1)+Y_1 \left(16 \beta _i+4 (-i+n-3)\right)-4 (i-1) (-i+n+1) Y_{i+1}\nonumber\\
& &-8 Y_{i+1} Z_{i-2}+8 Y_1
   Y_{i+1}+Z_{i-2} \left(4 (-i+n-3) (n-i)-16 \beta _i\right),
   \nonumber\\
\mathcal{Q}&=&8 i (n-4) \beta _i+4 i (n-i) Z_{i-2}+Y_1 \left(4 (4-i)-16 \beta _i\right)\nonumber\\
& &+\left(16 \beta
   _i-4 (i-4) (i-1)\right) Y_{i+1}-8 Y_1 Z_{i-2}+8 Y_{i+1} Z_{i-2},
   \nonumber\\
R&=& -4 \left(-i^2+4 \beta _i+(i-2) n+i+4\right)+8 Y_{i+1}+8 Z_{i-2}+8 Y_1.
\end{eqnarray*}
    
After lengthy computation and analysis, we can show that the Casimir operator $K_i$ can be expressed in terms of only the central elements
\begin{eqnarray}
K_i &=&16  \left(2 (n-4) \beta _i  (-i^2+(i-1) n+i+2 )+ (n^2-4 n ) \beta_i^2\right )\nonumber\\
& &-16 Y_1  \left(\beta _i  (n(2 i-1) -2  (i^2-i+2 ) )+i^2+4
   \beta _i^2+(3-i) n-i-8 \right)\nonumber\\
& &-8 Y_{i+1} Z_{i-2}  (-i^2+4 \beta _i+(i+1)   n+i-8 )\nonumber\\
& &+16 Z_{i-2} \left (\beta _i  (-(i+4) n+4 (i+2)+n^2 )+(i-n) (n-2
   (i+1)) \right)\nonumber\\
& &+16 Y_{i+1}  \left(\beta _i ((i-1) n-4 (i-3))+(i-1) (n-2 (i-2)) \right)\nonumber\\
& &+8Y_1 Z_{i-2}  (-4 \beta _i-i+n-8 )+4 (-i+n-4) (n-i) Z_{i-2}^2
\nonumber\\
& & +4 Y_1^2 (8 \beta _i-3 )-8 Y_1  (4 \beta _i-i+9 ) Y_{i+1}
+16 Y_1 Y_{i+1} Z_{i-2}\nonumber\\
& &-16 Y_{i+1} Z_{i-2}^2-16 Y_{i+1}^2 Z_{i-2}+4 (i-5) (i-1) Y_{i+1}^2.
   \label{Kib}
\end{eqnarray}  
This generic formula of the Casimir operator is obtained by first deriving the expressions for each sub-structure i with $2\leq i \leq n-1$ up to $n=9$, examining the pattern of the results and then proposing and proving the general formula for any $i\geq 2$, It can be seen that unlike the cases sub-structures i and 2, the Casimir operator $K_i$ for sub-structure $i$ with $i\geq 2$ do contain the cubic terms of the central elements $Y_ {1},Y_ {i+1},Z_ {i-2}$.

Inserting (\ref{Kib}) to (\ref{phi-i}) we find that the structure function $\Phi_i(\mathcal {N}_i,u_i)$ is factorized as 
\begin{eqnarray}
\Phi_i(\mathcal{N}_i,u_i) &=&\left(\mathcal{N}_i + u_i - \frac{1}{2}( m_{i-2} + v_i+1)\right)\left(\mathcal{N}_i + u_i - \frac{1}{2}( m_{i-2} - v_i+1)\right)\nonumber\\
& &\left(\mathcal{N}_i + u_i - \frac{1}{2}(- m_{i-2} + v_i+1)\right)\left(\mathcal{N}_i + u_i - \frac{1}{2}(- m_{i-2} - v_i+1)\right)
\nonumber\\
& &\left(\mathcal{N}_i + u_i - \frac{1}{2}(y_1 + y_{i+1}+1)\right)\left(\mathcal{N}_i + u_i - \frac{1}{2}(y_1 - y_{i+1}+1)\right)
\nonumber\\
& &\left(\mathcal{N}_i + u_i - \frac{1}{2}(-y_1 + y_{i+1}+1)\right)\left(\mathcal{N}_i + u_i - \frac{1}{2}(-y_1 - y_{i+1}+1)\right),
\end{eqnarray}
where $m_{i-2}$ , $v_{i}$, $y_{1}$ and $y_{i+1}$  satisfy 
\begin{eqnarray}
&&Z_{i-2}=\frac{(i-3)^2}{4}-m_{i-2}^2,  ~~~~~~
 v_i^2=\frac{1+8 \beta _i}{4}, \nonumber\\  
&&Y_{1}=\frac{(n-2)^2}{4}-y_{1}^2, ~~~~~~
Y_{i+1}=\frac{(n-2-i)^2}{4}-y_{i+1}^2. \label{4.18}
\end{eqnarray}
Note that when $i=2$, we have $m_0^2=\frac{1}{4}-Z_0=\frac{1+8\beta_1}{4}=v_1^2$.

\section{Algebraic derivation of the energy spectrum}

We first examine the finite-dimensional representation of the quadratic algebras through the realisation of them in terms of the deformed oscillators. Let $n_i$ be the eigenvalues of the number operators $\mathcal{N}_i$. Then obviously we have the boundary conditions for the structure functions at $n_i=0$,
\begin{equation}
\Phi_1(0,E,u_1)=0 
\end{equation}
\begin{equation}
\Phi_i(0,u_i)=0, \quad  i=2,\ldots n-1
\end{equation}
We then impose the constraint
\begin{equation}
\Phi_1(p+1,E,u_1)=0 
\end{equation}
in order to have finite-dimensional representation of the deformed oscillators. It turns out that these equations are sufficient for us to fix the values of the parameters $u_i$ and obtain the energy spectrum E of the superintegrable system.

An algebraic derivation of the energy spectrum of the model is given as follows. In the derivation $m_i$ will take the role of $Z_i$ , i.e. $m_i$ and $Z_i$ are related via
\begin{equation}
  {Z_i=\frac{(i-1)^2}{4}-m_i^2}.\label{5.4}
\end{equation}
First solving $\Phi_1(0,E,u_1)=0$, we get the possible value of constant $u_1$,
\begin{equation}
  u_1=\frac{1}{2}+m_{n-2}\label{5.5},
\end{equation}
where $m_{n-2}$ satisfies the relation corresponding to $i=n-2$ in (\ref{5.4}).
Inserting (\ref{5.5}) and solving $\Phi_1(p+1,E,u_1)=0$, we get $E$ in terms of $m_{n-2}$,
\begin{equation}
  E=-\frac{2\gamma^2}{\left(2+2p+2m_{n-2}\right){}^2}.\label{5.7}
\end{equation}
It follows that the energy depends on $m_{n-2}$ related to other sub-structures. Thus to obtain the energy spectrum we need to determine the value of $m_{n-2}$ from the other sub-structures.

Solving $\Phi_i(0,u_i)=0 $, we get the possible values of constant $u_i$
\begin{equation}
  u_i=\frac{1}{2}( m_{i-2}+v_i+1).\label{5.8}
\end{equation}
From the realization of the sub-structure $i$ we get $Z_{i-1}$ in terms of $u_i$,
\begin{equation}
  Z_{i-1}=-4 \left(\left(n_i+u_i\right){}^2-\frac{1}{16} (i-2)^2\right).\label{5.9}
\end{equation}
Comparing with (\ref{5.4}) gives 
\begin{equation}
  m_{i-1}=2\left(n_i+u_i\right). \label{5.11}
\end{equation}
Substituting (\ref{5.8}) into (\ref{5.11}), we obtain the recursive relation,
\begin{equation}
 m_{i-1}= m_{i-2}+ 2n_i +v_i+1\label{5.12}
\end{equation}
This recursive relation will be used to obtain the value of $m_{n-2}$.

The initial value $m_1$ can be determined from the sub-structure 2. To see this we first solve $\Phi_2(0,u_2)=0$ to obtain the possible value of the constant $u_2$
\begin{equation}
 u_2=\frac{1}{2}(v_1+v_2+1).\label{5.13}
\end{equation}
Then from  (\ref{5.9}) and (\ref{5.4}), we find
\begin{equation}
  m_1=2\left(n_2+u_2\right). \label{5.16}
\end{equation}
which, by means of (\ref{5.13}), becomes
\begin{equation}
  m_1=2n_2+v_1+v_2+1. \label{5.17}
\end{equation}

Now from the recursive relation (\ref{5.12}) and the initial value (\ref{5.17}) $m_1$ we get the expression for $m_{n-2}$
\begin{equation}
  m_{n-2}=2\sum_{j=2}^{n-1} n_j+\sum_{i=1}^{n-1} v_i+(n-2). \label{5.18}   
\end{equation}
Substituting (\ref{5.18}) into(\ref{5.7}) yields the energy spectrum $E$ of the system,
\begin{equation}
 E=-\frac{ \gamma{}^2}{2\left(p+2\sum_{j=2}^{n-1} n_j+ \sum_{i=1}^{n-1} v_i+(n-1)\right){}^2},\label{finalE}
\end{equation}
where $v_i$ are given in (\ref{4.18}).

To compare our result with the one obtained in \cite{model} using the separation of variable method, we note that $p_i$ in \cite{model} are related to $\beta _i$ by
\begin{equation}
  {{\beta}_{i}=\frac{1}{2}p_i(p_i-1)}.
\end{equation}
Thus in terms of $p_i$, $v_i$ can be written as
\begin{equation}
  {v_i=p_i-\frac{1}{2}}.\label{5.21}
\end{equation}
Inserting (\ref{5.21}) into (\ref{finalE}), we have
\begin{equation}
 E=-\frac{ \gamma{}^2}{2\left(p+2\sum_{j=2}^{n-1} n_j+ \sum_{i=1}^{n-1} p_i+\frac{n-1}{2}\right)^2}.
\end{equation}
This expression coincides with (\ref{2.2}) obtained in \cite{model} via separation of variables if we make the identifications
\begin{equation}
  {n_i=J_{i-1}}, ~~~~~~ p= {N}_1+{N}_2.
\end{equation}

\section{Conclusions}

We have constructed using a direct approach and integrals of motion a algebraic structure of a family of $n$-dimensional quantum superintegrable system proposed in \cite{model}. We have found a chain of quadratic algebra structures for the system and identified the elements of each sub-structures. We have presented the commutation relations satisfied by elements of each sub-structure. We have found the Casimir operators corresponding to each sub-quadratic algebra. Realization of each quadratic algebra in terms of the deformed oscillators has been presented by obtaining each of the Casimir in terms of the Central element which belong to a set of commuting operators. Utilizing the underlying symmetry algebra, algebraic derivation of the Energy spectrum for the system has been performed. The sub-structures discovered will allow further application of algebraic method for higher rank polynomial algebras and corresponding superintegrable models. Quadratic deformation of Lie super algebras have been observed in other context \cite{exa}.

\section*{Acknowledgements}
IM was supported by Australian Research Council Discovery Project DP 160101376. YZZ was supported by
National Natural Science Foundation of China (Grant No. 11775177).

\paragraph{}

% BIBLIOGRAPHY
% use BIBTEX if you want
%\bibliographystyle{JHEP}
%\bibliography{yourBIBfiles}


\end{document}